# Regional Flood Risk Projections under Climate Change

Sanjib Sharma[a,b], Michael Gomez[b], Klaus Keller[a,c], Robert Nicholas[a], Alfonso Mejia[b*]

[a]Earth and Environmental Systems Institute, The Pennsylvania State University, University Park, Pennsylvania

[b]Department of Civil and Environmental Engineering, The Pennsylvania State University, University Park, Pennsylvania

[c] Department of Geosciences, The Pennsylvania State University, University Park, Pennsylvania

*Corresponding author: Alfonso Mejia; aim127@psu.edu

**Key Points**

- We develop a framework to estimate regional riverine flood inundation hazards and exposure for future climatic conditions.
- We find that assuming a stationary climate can underestimate flood hazards.
- The analysis requires an integrated approach since the uncertainty in flood inundation projections is impacted by uncertainties surrounding climate change, hydrology, and hydraulics.
- The effects of the considered climate uncertainty on projected flood hazards dominate the effects of hydrologic and hydraulic uncertainties.

## ABSTRACT

Flood-related risks to people and property are expected to increase in the future due to environmental and demographic changes. It is important to quantify and effectively communicate flood hazards and exposure to inform the design and implementation of flood risk management strategies. Here we develop an integrated modeling framework to assess projected changes in regional riverine flood inundation risks. The framework samples climate model outputs to force a hydrologic model and generate streamflow projections. Together with a statistical and hydraulic model, we use the projected streamflow to map the uncertainty of flood inundation projections for extreme flood events. We implement the framework for rivers across the state of Pennsylvania, United States. Our projections suggest that flood hazards and exposure across Pennsylvania are overall increasing with future climate change. Specific regions, including the main stem Susquehanna River, lower portion of the Allegheny basin and central portion of Delaware River basin, demonstrate higher flood inundation risks. In our analysis, the climate uncertainty dominates the overall uncertainty surrounding the flood inundation projection chain. The combined hydrologic and hydraulic uncertainties can account for as much as 37% of the total uncertainty. We discuss how this framework can provide regional and dynamic flood-risk assessments and help to inform the design of risk-management strategies.

## 1. Introduction

Floods drive major damages to communities across the globe, with an estimated annual average loss of US $104 billion (UNISDR 2015). These impacts are expected to rise in the future as the climate is changing and urbanization is changing exposure (Alfieri et al. 2015; Hirabayashi et al. 2013; Wing et al. 2018; Winsemius et al. 2016). There is increasing interest in improving the understanding and quantification of future flood hazards in a changing climate (Judi et al. 2018; Gangrade et al. 2020; FSF 2020). There is also growing recognition of the critical need to enhance

the communication of flood hazards and its associated uncertainties to inform stakeholders and the design and implementation of flood risk management strategies (Collet et al. 2018; Kundzewicz et al. 2018; Sanders et al. 2020). The design of flood-risk management strategies can be improved by information about future flood hazards that is based on the best current understanding of the future climate system (Alfonso et al. 2016; Sorribas et al. 2016; Srikrishnan et al. 2019; Zarekarizi et al. 2020).

In the United States (US), the Federal Emergency Management Agency (FEMA) is responsible for producing Flood Insurance Rate Maps (FIRMs) specifying the boundaries of zones deemed to be most vulnerable to fluvial and coastal flooding (FEMA 2019). The most common delineations are the 100-yr and 500-yr floodplain boundary, which depict the inundation extent for floods assessed to have a 1% and 0.2%, respectively, chance of being reached or exceeded in any given year. These estimates are widely used by federal, state, and local agencies in the US for setting national-level flood insurance requirements and managing development in the floodplain at the community level (FEMA 2019).

The FEMA flood maps can provide very useful information, but they are also subject to several limitations (Bales and Wagner 2009; Merwade et al. 2006; 2008; Pappenberger et al. 2005, 2006). For one, they do not account for uncertainties associated with the estimation and mapping of flood inundation (Merwade et al. 2008; Lin et al. 2013) and assume that the flood-event time series is stationary (Villarini et al. 2009). Historical records often reveal nonstationarity due to multiple factors such as changes in precipitation patterns, land cover alterations, and water transfers, among others (Jovanovic et al. 2016, 2018). Here we design, test, and discuss a framework to derive probabilistic flood inundation projections that accounts for key nonstationarities associated with climate change.

Flood risk can be assessed in terms of hazard, exposure, and vulnerability (Arnell and Gosling 2014; Hirabayashi et al. 2013; Ruhi et al. 2018; Wing et al. 2018; Keller et al. 2020). Hazard refers to the nature, magnitude and probability of a flood event. Exposure in this case characterizes factors such as the population and value of assets within the floodplain that are likely to experience flooding. Vulnerability characterizes how sensitive the impacts are for a given hazard and exposure. A sound understanding of these risk drivers is important to inform the design of risk management strategies, for example, deciding when and where to acquire land, or floodproofing infrastructure. Past flood risk assessment have focused on the global scale (Alfieri et al. 2017; Hirabayashi et al. 2013; Jongman et al. 2015; Winsemius et al. 2016; Ward et al. 2016), continental scale (Alfieri et al. 2015; Feyen et al. 2012; Wing et al. 2018) or local watershed scale (Cheng et al. 2017; Dibike and Coulibaly 2005; Dobler et al. 2012; Wu et al. 2014). Recently, Wing et al. (2018) presented new estimates of current and future flood risk across the US. This study finds that nearly 41 million of the US population live within the 1% annual exceedance probability floodplain (compared to only 13 million when calculated using the FEMA flood maps). However, most hydrological responses to extreme events are highly regional, and water resources management is often carried out at river basin scales (Hattermann et al. 2017; Varis et al. 2004). The local decision-making scales requires local information (Aich et al. 2016; Collet et al. 2018; Demaria et al. 2016; Huang et al. 2013; Moel et al. 2015).

The main objective of this study is to develop an integrated hydroclimatic modeling framework to assess regional flood hazards and exposure for future climatic conditions. The framework combines five main components: i) statistically downscaled daily precipitation and near-surface temperature projections (Pierce et al. 2014); ii) the National Oceanic and Atmospheric Administration's (NOAA) Hydrology Laboratory-Research Distributed Hydrologic Model (HL-RDHM) (Koren et al. 2004); iii) the Generalized Pareto distribution (GPD)-based extreme value model (Coles 2001); iv) the LISFLOOD-FP hydraulic model (Neal et al. 2012); and v) a uncertainty quantification methods applied to these components. The framework involves hydrologic modeling driven by climate ensembles to generate streamflow projections; statistical modeling to estimate flood peaks; hydraulic modeling to estimate water surface elevations; terrain analysis to map the inundation extents; and uncertainty decomposition to quantify the key uncertainty sources in flood inundation projections. We implement this integrated hydroclimatic modeling framework for a regional scale case-study to analyze two interrelated questions: i) How does climate change alter regional flood hazard and exposure projections? and ii) What drives the total uncertainty in flood inundation projections?

Flood inundation projections are subject to multiple sources of uncertainty. These sources include projections of future climate, as well as hydrologic and hydraulic modeling. Previous studies highlight the dominating influence of climate model-driven uncertainty relative to the total uncertainty in extreme flood projections (Hirabayashi et al. 2013; Vetter et al. 2017). Most studies focus on quantifying individual sources uncertainties in extreme flood projections (Bosshard et al. 2013; Qi et al. 2016), and are silent on the relation between individual sources of uncertainty and the total uncertainty of flood inundation projections (Kim et al. 2019). In addition, previous studies often ignore the effects of hydraulic modeling and mapping on flood inundation extents when assessing climate change impacts on flood projections (Bosshard et al. 2013; Kim et al. 2019). Thus, the relative contribution of hydraulic uncertainty to the total flood inundation projection uncertainty is often unknown. Here we quantify key uncertainties in the flood inundation projection chain, including uncertainties in climate model outputs, hydrologic and hydraulic modeling for a case study.

We implement an integrated hydroclimatic modeling framework to assess flood hazards and exposure across the state of Pennsylvania, which is located in the Mid-Atlantic region of the US. Flooding has been the most frequent and most damaging natural disaster in Pennsylvania (PEMA 2017), causing an average $92 million loss a year during 1996 and 2014 (USDOE 2015). Regionally, there is strong evidence of rising temperatures, altered precipitation patterns (Hayhoe et al. 2018; Shortle et al. 2015) and increasing intensity of flood events (Sagarika et al. 2014; Zhang et al. 2010). Historical climate records show that temperatures in Pennsylvania have already increased by more than 1.8 °F since the early 20th century (Shortle et al. 2015). Similarly, average annual precipitation in Pennsylvania has increased by approximately 10% over the past 100 years and, by 2050, it is expected to increase by 8% (Kang and Sridhar 2018; Shortle et al. 2015). There has been an increase in the frequency and magnitude of damaging flood occurrence in communities across Pennsylvania. For instance, the most destructive floods in the Susquehanna and Delaware River basins have occurred in recent years, each associated with different flood-generating mechanisms: Hurricane Ivan (September 2004), late winter–early spring extratropical systems (April 2005), warm-season convective systems (June 2006), and tropical storm Lee (September 2011) (Armstrong et al. 2014; Smith et al. 2010, 2011). There is a critical need to better understand future flood risks in Pennsylvania (Iulo et al. 2020).

The remainder of the paper is organized as follows. Section 2 presents the methods used in this study. The main results and their implications are examined in Section 3. Lastly, Section 4 summarizes key findings.

## 2. Materials and Methods

We develop and implement an integrated hydroclimatic modeling framework consisting of downscaled climate model outputs, a hydrologic model, a nonstationary extreme value model, a hydraulic model, and uncertainty quantification (Fig.1). The proposed framework uses downscaled climate projections for precipitation and temperature to force the NOAA's HL-RDHM (Koren et al. 2004) resulting in future projected streamflow at selected gaged locations in Pennsylvania. The projected streamflow is statistically postprocessed to compute extreme flows with 100-yr return periods. The 100-yr flood peaks are used as boundary conditions to the hydraulic model LISFLOOD-FP (Bates et al. 2013) to generate flood inundation projections. Finally, the flood inundation projections are used to analyze flood hazards and exposure. Next, we describe the datasets, models and techniques involved in overall workflow (Fig. 1).

*a. Datasets*

For the observational datasets, we use multi-sensor precipitation estimates (MPEs), gridded near-surface air temperature, and streamflow observations at selected US Geological Survey (USGS) gages. These observational datasets are used to calibrate the hydrological model. We use hourly gridded MPEs and near-surface air temperature data at 4 x 4 $km^2$ from the NOAA's Middle Atlantic River Forecast Center. Similar to the National Center for Environmental Prediction's stage-IV dataset (Prat and Nelson 2015), the MPEs provided by the River Forecast Center represent a continuous time series of hourly, gridded precipitation observations at $4 \times 4$ $km^2$ cells. The gridded near-surface air temperature data at $4 \times 4$ $km^2$ resolution were developed by the River Forecast Center by combining multiple temperature observation networks as described by Siddique and Mejia (2017). We obtained daily streamflow observations for the selected gaged locations from the USGS (https://waterdata.usgs.gov/nwis/rt).

For the climate ensemble, we use the Localized Constructed Analogs (LOCA; Pierce et al. 2014) daily precipitation and near-surface temperature outputs for 13 general circulation models from the Coupled Model Intercomparison Project CMIP5 datasets. The CMIP5 outputs are for model runs forced by emissions corresponding to Representative Concentration Pathway scenario 8.5 (RCP8.5) (Taylor et al. 2012; Meinshausen et al. 2011). The 13 considered climate models are CESM1-CAM5, CanESM2, EC-EARTH, GFDL-ESM2M, GISS-E2-R, HadGEM2-CC, HadGEM2-ES, IPSL-CM5A-LR, IPSL-CM5A-MR, MPI-ESM-LR, MPI-ESM-MR, BCC-CSM1-1, and INM-CM4 (Table 1). These climate models are selected since they demonstrate improved skill and reliability across the Northeast region (Demaria et al. 2016). The downscaled model outputs are at 6 x 6 km$^2$ spatial resolution and provided at https://gdo-dcp.ucllnl.org/downscaled_cmip_projections/dcpInterface.html. We use bilinear interpolation to transform the climate model outputs to the 4 × 4 km$^2$ grid cell resolution of the hydrological model.

*b. Hydrological model*

We use the NOAA's HL-RDHM as the hydrological model (Koren et al. 2004). HL-RDHM is a spatially distributed conceptual model, where the basin system is divided into regularly spaced, square grid cells to sample spatial heterogeneity and variability. We run HL-RDHM at a spatial resolution of 4 x 4 km$^2$.

Within HL-RDHM, we use the Sacramento Soil Moisture Accounting model with Heat Transfer (SAC-HT) to represent hillslope runoff generation, and the SNOW-17 submodel is used to represent snow accumulation and melting. The hillslope runoff, generated at each grid cell by SAC-HT and SNOW-17, is routed to the stream network using a nonlinear kinematic wave algorithm (Koren et al. 2004; Smith et al. 2012). Likewise, we route flows in the stream network downstream using a similar nonlinear kinematic wave algorithm that accounts for parameterized stream cross-section shapes (Smith et al. 2012; Koren et al. 2004). HL-RDHM has successfully been used before to predict and map flood events under a wide range of conditions within our study area (Gomez et al. 2019; Sharma et al. 2018; Siddique and Mejia 2017; Zarzar et al. 2018).

Following previous studies (Koren et al. 2004; Reed et al. 2004), we calibrate HL-RHDM by adjusting a subset of model parameters. Specifically, we select ten out of the 17 SAC-HT parameters based upon prior experience and preliminary parameter sensitivity tests. The adjusted parameters are associated with baseflow, percolation, evaporation, storm runoff and channel routing. To calibrate the selected HL-RDHM parameters, we multiply each a-priori parameter field by a factor. It is these multiplying factors that are actually calibrated, not the parameter values in each grid cell. The multiplying factors are adjusted manually first; once the manual changes do not yield noticeable improvements in model performance, the factors are optimized using the automatic technique of stepwise line search (SLS) (Kuzmin 2009; Kuzmin et al. 2008). We use this approach as it is readily available within HL-RDHM and has been shown to provide reliable parameter estimates (Kuzmin 2009; Kuzmin et al. 2008; Sharma et al. 2018; Siddique and Mejia 2017). We optimize the objective function, *OF*,

$$OF = \sqrt{\sum_{i=1}^{m} [q_i - s_i(\Omega)]^2}, \qquad (1)$$

where $q_i$ and $s_i$ denote the daily observed and simulated flows at time *i*, respectively; $\Omega$ is the parameter vector being estimated; and *m* is the total number of days used for calibration.

To assess the performance of HL-RDHM, we compare the daily simulated flows with daily observed flows from USGS gauges. We obtain the simulated flows by forcing the model with gridded precipitation and near-surface temperature observations. The model is calibrated during 2008-2012 at selected USGS gauge stations for the major river basins in Pennsylvania. We use the year 2007 to spin up the model. To assess the quality of the streamflow simulations from HL-RDHM, we employ the modified correlation coefficient ($R_m$) (McCuen and Snyder 1975) and Nash-Sutcliffe Efficiency (*NSE*) (Nash and Sutcliffe 1970) as performance measures. Overall, the performance of the simulated streamflow is arguably satisfactory at both the calibration and validation gauge stations (Fig. 2). The $R_m$ for the calibration stations is greater than 0.90. Likewise, the *NSE* ranges from 0.75 to 0.85 for the calibration stations (Fig. 2). The $R_m$ for the validation stations ranges from ~0.80 to 0.95, while the *NSE* ranges from ~0.60 to 0.75 (Fig. 2).

*c. Nonstationary extreme value analysis of flood events*

Statistical distributions of extreme floods are widely used to inform flood-sensitive infrastructure design, floodplain mapping, risk assessment and environmental management (Bracken et al. 2018; Villarini et al. 2009). Traditional approaches for extreme flood estimation use the historical records and assume stationarity in streamflow time series (FEMA 2019). In a changing climate, however, traditional approaches can lead to poor hazard estimates (Pralle 2019). Here we compute extreme floods allowing for nonstationary conditions. For this, we use the peak over threshold approach with a Poisson point process parameterization of the Generalized Pareto distribution (GPD) (Coles 2001). As opposed to selecting a single peak flood per year, the peak over threshold approach enables us to use daily streamflow data that exceed a specified threshold to fit the GPD model parameters. This approach allows us to consider more information about the extremes. We use a constant threshold $\mu$ equal to the 95th percentile of the daily maximum flow (Lang et al. 1999).

The probability density function for the GPD model is defined by:

$$f(x(t); \mu(t), \sigma(t), \xi(t)) = \frac{1}{\sigma(t)}\left(1 + \xi(t)\frac{x(t) - \mu(t)}{\sigma(t)}\right)^{-\left(\frac{1}{\xi(t)}+1\right)}, \quad (2)$$

where $x$ is the daily streamflow observations for current flood peak estimates or the projected daily streamflow for future flood peaks, $\sigma$ is the scale parameter that governs the width of the distribution, and $\xi$ is the shape parameter that governs the heaviness of the distribution's tail, all as functions of time $t$. The Poisson process governs the probability $p$ of observing $n$ exceedances of threshold $\mu$ during a time interval $\Delta t$:

$$p(n(t); \lambda(t)) = \frac{(\lambda(t)\Delta t)^{n(t)}}{n(t)!} exp(-\lambda(t)\Delta t), \quad (3)$$

where $\lambda$ is the Poisson rate parameter. We incorporate potential nonstationary into the GPD model by allowing the model parameters to covary with time (Papoulis and Pillai 2002; Koutsoyiannis 2006) such that

$$\lambda(t) = \lambda_0 + \lambda_1(t),$$
$$\sigma(t) = exp[\sigma_0 + \sigma_1(t)], \quad (4)$$
$$\xi(t) = \xi_0 + \xi_1(t).$$

$\lambda_0, \lambda_1, \sigma_0, \sigma_1, \xi_0 \text{ and } \xi_1$ are regression parameters.

We adopt a Bayesian approach and use Markov chain Monte Carlo sampling for parameter estimation (Stephenson and Tawn 2004). This approach combines the knowledge brought by a prior distribution and the observations into the parameters' posterior distribution. We produce $10^5$ iterations for one MCMC chain and remove the first 25,000 iterations to account for burn-in. We develop separate GPD models for the period of 1981-2017 and 2020-2099 to estimate 100-yr flood peaks in 2017 and 2099, respectively. The GPD model for the period of 1981-2017 is based on historical streamflow observations, whereas the GPD model for the period of 2020-2099 is based on flow projections obtained by forcing HL-RDHM with the GCM outputs.

*d. Regionalization of the 100-yr flood peaks*

To map the flood extent for an entire region with the LISFLOOD-FP hydraulic model, we need to estimate the 100-yr flood peak for any river location within that region since these estimates are used as boundary conditions by the LISFLOOD-FP. From the flood frequency analysis with the GPD distribution, we obtain estimates of the 100-yr flood peak at selected gaged locations. To regionalize these estimates, we use a scaling relationship between the 100-yr peak $Q_p$ and the drainage area $A$:

$$Q_p = \beta A^\alpha, \tag{5}$$

where $\beta$ is the proportionality constant and $\alpha$ the scaling exponent. The parameters in equation (5) are estimated using ordinary least squares (England et al. 2018; Smith et al. 2011).

*e. Flood inundation mapping*

We use the LISFLOOD-FP hydraulic model with the sub-grid formulation of Neal et al. (2012) to simulate and project flood inundations along rivers in Pennsylvania. LISFLOOD-FP (Bates et al. 2013) is a 2-D hydraulic model for subcritical flow that solves the local inertial form of the shallow water equations using a finite difference method on a staggered grid. The model requires as input ground elevation data describing the floodplain topography, channel bathymetry information (river width, depth and shape), and inflow to the modeling domain as the boundary condition information. We use floodplain topography information from the Pennsylvania Spatial Data Access archive (PASDA, https://www.pasda.psu.edu/), and extreme flood events from the nonstationary GPD model and scaling relationship in equation (5) .

We run LISFLOOD-FP for all rivers in Pennsylvania using a 30-meter Digital Elevation Model (DEM) from PASDA. In addition, we use other DEM resolutions, 1- and 10-meter, at selected locations as part of our uncertainty analysis (Subsection 2g). Peak flows with 100-yr return period are used as the inflow boundary condition. Channel bathymetry is estimated using regional, hydraulic-geometry scaling relationships from previous studies (Cinotto 2003; Chaplin 2005; Clune et al. 2018; Roland and Hoffman 2011). We use a constant Manning's roughness value of 0.045 for both the channel and floodplain. The chosen Manning's roughness value is similar to values previously used for rivers in Pennsylvania (Newlin and Hayes 2015). Structural flood mitigation barriers, such as flood walls, levees and dikes, are not explicitly incorporated into the hydraulic analysis.

*f. Flood hazards and exposure analysis*

We assess flood hazards and exposure in all cities and boroughs in Pennsylvania for current and projected future climatic conditions. Pennsylvania has 959 boroughs and 56 cities. Boroughs are defined as incorporated political subdivisions and are mostly less populous than cities. Most of the boroughs in Pennsylvania have populations under 5,000, though there are some exceptions. Some major cities in Pennsylvania include Philadelphia (with more than one million residents), Pittsburgh, and Scranton.

We characterize flood hazards by the extent of flooding resulting from the 100-yr flood peak. We calculate flood exposure from the population within the flood extent. Flood hazards and exposure are both expressed as percentages. The percent hazard is the flood inundation area standardized to the total area of a city/borough. The percent exposure is the population in the flood inundation area standardized to the total population of a city/borough. We use population from the 2010 US Census Bureau data (US Census Bureau 2010). For a particular borough/city, we disaggregate the total population based on the urban development intensity of each developed grid cell in the 2011 National Land Cover Database (Homer et al. 2012). The National Land Cover Database provides the definitive Landsat-based, 30-meter resolution, land cover database for the US.

*g. Assessing key uncertainties of the flood inundation projections*

To assess key uncertainties in flood inundation projections, we use a cumulative uncertainty approach (Kim et al. 2019). This approach decomposes the total uncertainty to individual uncertainty sources, such that the sum of the uncertainties from individual sources is equal to the total uncertainty in flood inundation projections. We consider uncertainty from three key sources in the flood inundation projection chain: climate, hydrology, and hydraulics. Climate uncertainty in this case is the uncertainty sampled by the information derived from the global climate model runs. Hydrologic uncertainty refers to the uncertainty in the GPD model parameters obtained using Bayesian inference. Hydraulic uncertainty is sampled by the topographic uncertainty due to the choice of different DEM resolutions: 1-, 10-, and 30-meter. For the uncertainty assessment, we compute the stage uncertainty as the contribution to uncertainty from each of the three stages (climate, hydrology and hydraulic) in the flood inundation projection chain. Note that this analysis samples a relatively small subset of potentially important uncertainties. For example, we sample just a subset of climate models, initial conditions, and climate model parameters (Sriver et al. 2015). Further, the analysis accounts for the hydrologic uncertainty associated with the GPD's parameters while the HL-RDHM model's parameters are fixed.

To characterize the stage uncertainty, we first compute the conditional cumulative uncertainty up to a particular stage, which is defined as the variation in the projected flood inundation extent due to the modeling choices up to that stage, while the choices beyond that stage are fixed. For instance, conditional cumulative uncertainty up to the hydrology stage represents the variation in flood inundation projections due to the choice of different GCM outputs and GPD parameters, while the DEM resolution is fixed. Then the marginal cumulative uncertainty up to a particular stage is computed as an average of conditional cumulative uncertainties. Finally, we compute the uncertainty of each stage as the difference between successive marginal cumulative uncertainties.

Specifically, we calculate stage uncertainty as follows. We denote by $K$ the total number of stages in the flood inundation projection chain, in this case $K=3$ (climate, hydrology and hydraulic stage). For a particular stage $k$, there are $\chi_k$ models/scenarios. The cumulative uncertainty up to stage $k$ is defined as the variation in the projected values due to the choice of models/scenarios up to stage $k$, while the models/scenarios after stage $k$ are fixed. The cumulative uncertainty up to stage $k$ is denoted by $U^{cum}(\chi_1,...,\chi_k)$. For the specific models/scenarios of stage $k$, we let $P(x_1, x_2, ..., x_K)$ be the projected value using the models/scenarios $x_k, k = 1, ..., K$. For a given model/scenario after stage $k$, the set of projected values are:

$$q_{x_{k+1},...,x_K} = \{P(x_1,...,x_k,x_{k+1},...,x_K): x_j \in \chi_j, j = 1,...,k\}. \quad (6)$$

Then $U^{cum}(q_{x_{k+1},...,x_K})$ can be interpreted as the conditional cumulative uncertainty up to stage $k$ while the models/scenarios after stage $k$ are fixed as $x_{k+1,...,}x_K$. The marginal cumulative uncertainty up to stage $k$ is the average of conditional cumulative uncertainties defined as:

$$U^{cum}(\chi_1,...,\chi_k) = \frac{1}{\prod_{j=k+1}^{K} n_j} \sum_{x_{k+1} \in \chi_{k+1}} ... \sum_{x_K \in \chi_K} U(q_{x_{k+1},...,x_k}). \quad (7)$$

Since the cumulative uncertainty increases monotonically (Kim et al. 2019), we can define the uncertainty of each stage as the difference between successive cumulative uncertainties. That is, the uncertainty of stage $k$, denoted by $U^{cum}(\chi_k)$, is defined as:

$$U^{cum}(\chi_k) = U^{cum}(\chi_1, \ldots, \chi_k) - U^{cum}(\chi_1, \ldots, \chi_{k-1}). \tag{8}$$

The uncertainty of each stage is the amount of contribution to the cumulative uncertainty. Also, the sum of uncertainties of individual stages is always equal to the total uncertainty $U^{cum}(\chi_1, \ldots, \chi_k)$. The uncertainty of stage $k$ can also be defined as the sum of the variation of the main effect of stage $k$ and the variations of the interactions between stage $k$ and stages after $k$.

The stage uncertainty and cumulative uncertainty are both expressed in terms of the range (Chen et al. 2011) and standard deviation (Bosshard et al. 2013) of the projected flood inundation extents. For a real number $y = q_{x_{k+1},\ldots,x_K}$ and a set of $y = \{y_1, \ldots, y_n\}$, the range and standard deviation are defined as:

$$\text{Range} = max_{1 \leq i \leq n} y_i - min_{1 \leq i \leq n} y_i, \tag{9}$$

$$\text{Standard deviation} = \frac{1}{n} \sum_{i=1}^{n}(y_i - \bar{y}), \tag{10}$$

where $\bar{y} = \frac{1}{n}\sum_{i=1}^{n} y_i$.

## 3. Results and Discussion

*a. Regional flood frequency analysis*

We summarize the results obtained from the regional flood frequency analysis using the scaling relationship between the 100-yr flood peak $Q_p$ and the drainage area $A$. For years 2017 and 2099, we find that the scaling of the 100-yr flood peaks with drainage area performs reasonably well with a Pearson correlation coefficient $R$ exceeding 0.9 (Fig. 3). The years 2017 and 2099 are used for all our flood risk analyses and results. Fig. 3 shows the mean 100-yr flood peak estimates and projections in years 2017 and 2099, respectively, as well as the estimates under the stationary assumption for historical flood records. Flood peak estimates and projections in years 2017 and 2099 are based on the estimated and projected streamflow, respectively, obtained by forcing HL-RDHM with the climate models' outputs.

The nature of the scaling relationships (Fig. 3) has implications for flood-risk management. Based on the power-law fits in Fig. 3, the ratio between the 100-yr flood peak for 2099, $Q_{p,2099}$, and 2017, $Q_{p,2017}$, is $Q_{p,2099}/Q_{p,2017} = 1.50 A^{-0.0196}$. The ratio of the flood peaks indicates that the drainage area has a small and slightly negative amplification effect on future flood peaks since the value of the scaling exponent for the ratio, -0.0196, is small. The ratio of the flood peaks also shows that the flood-peak amplification associated with climate projections is mainly due to the value of the proportionality constants in the power-law fits. Taken together, these results suggest that future increases in flood peaks from climate projections tend to have a disproportionate effect on smaller basins. Many flood-related management decisions, for example, those associated with stormwater management, are made in smaller basins, which adds urgency to the need of adapting local flood regulations and design standards to account for potentially changing climate conditions.

The stationary assumption underestimates the 100-yr flood peaks relative to the nonstationary flood peaks (Fig. 3). In other words: assuming stationary flood peaks can underestimate actual flood hazards in our analysis. For example, the flood hazards further increase with the nonstationary flood peak projections for the year 2099 relative to the 2017 nonstationary flood estimates (Fig. 3). As compared to the flood peak estimates for 2017, the projected flood peak is higher, in terms of absolute values, in the case of larger basins, as implied by the scaling relationships obtained.

*b. Flood hazard and exposure from future climatic conditions*

Our results suggest that most cities (Fig. 4a-b) and boroughs (Fig. 4c-d) in Pennsylvania are projected to face higher flood hazards and exposure under future climatic conditions in the year 2099. Populous cities, however, seem to have relatively lower flood hazards and exposure. For instance, the three largest cities (Philadelphia, Pittsburgh and Scranton) show relatively low projected flood hazards and exposure compared to smaller cities (Fig. 4a-b). Our analysis suggests that these cities have less than 25% projected flood hazards and exposure in 2099.

Some of the cities with the highest projected flood hazards include Sunbury, Williamsport, Lock Haven, Warren, Bradford, Wilkes Barre, Johnstown, York and Connellsville (Fig. 4a). These cities are projected to have more than 50% flood hazards and exposure. The top three cities with the highest projected flood hazards are Lock Haven, Williamsport and Sunbury (Fig 4a), which are all located along the West Branch Susquehanna River. Importantly, based on NOAA's National Centers for Environmental Information (NCEI) storm events database (https://www.ncdc.noaa.gov/stormevents/), these three cities have all experienced a substantial number of flood events over time, including events within the last decade, suggesting that our flood hazard metric can capture hotspot areas prone to flooding within Pennsylvania. Specifically, the West Branch Susquehanna River at Williamsport has experienced 40 flood events between 1864 and 2018 (NCEI, 2020). Lock Haven has experienced 21 flood events between 1889 and 2010, while Sunbury has experienced 14 flood events between 1936 and 2011 (NCEI, 2020). The cities with the highest projected exposure include Bradford, Sunbury and Bethlehem (Fig. 4b), which are also cities with high flood hazards. Of all the considered cities, Sunbury demonstrates the highest projected flood hazards and exposure (Figs. 4a and 4b). Sunbury is located at the confluence between the West and North Branch of the Susquehanna River. This makes Sunbury particularly vulnerable to flooding as flood events can occur in either branch of the Susquehanna River (FCSMA 2020).

Boroughs around the high hazards/exposure cities tend to have higher hazards and exposure (Figure 4c-d). For example, most of the boroughs around the cities of Lock Haven, Williamsport and Sunbury, and particularly along the West Branch Susquehanna River, show higher flood hazards and exposure. Boroughs along the West Branch Susquehanna River, including Duboistown, Montgomery, Muncy, South Williamsport, Lewisburg, Milton and Selinsgrove are projected to have more than 50% flood hazards and exposure. Neighboring boroughs to the largest metropolitan area (Philadelphia) generally have lower flood hazards and exposure. For example, most of the boroughs around Philadelphia are projected to have less than 25% flood hazards and exposure.

Some of the projected flood hazards and exposure tend to be spatially clustered (Fig. 4). By visually inspecting our flood hazards and exposure maps (Fig. 4), we identify the most salient clusters. The high hazards and exposure clusters are primarily concentrated along the West Branch of the Susquehanna River, the southwest portion of the Susquehanna basin (also called Juniata sub-basin), the lower portion of the Allegheny basin (Fig. 2 and Fig. 4), and the central portion of Delaware River basin. The historical records associated with these clusters show that these areas are prone to frequent and severe flooding. For example, the mainstem of the Susquehanna and Delaware River have experienced several devastating floods in past decades, with the four most recent events in the years 2004, 2005, 2006 and 2011 (Gitro et al. 2014; Suro et al. 2009). These historical events caused record to near-record flood crests along most of the streams and rivers throughout the main stem Susquehanna and Delaware River. For instance, for the Susquehanna and Delaware River flood event of 2004, peak discharges on the unregulated reaches remained below a 50-year recurrence interval, while at the regulated reaches it equaled or exceed the estimated 100-year recurrence interval (Brooks, 2005). Many communities that were flooded during the 2004, 2005 and 2006 floods were again flooded in 2011 (Gitro et al. 2014).

Several factors, including development, stormwater management, floodplain encroachment and reservoir management, have been hypothesized as contributing and exacerbating factors for recent flooding in the Susquehanna and Delaware River basins (Brooks 2005; DRBC 2006). Many of these historical flood events have been caused by different flood generating mechanisms: Hurricane Ivan (September 2004); late winter–early spring extratropical systems (April 2005); warm-season convective systems (June 2006); and tropical storm lee (September 2011) (Smith et al. 2010, 2011). The importance of tropical cyclone projections points to an important caveat as well as a research need. This is because tropical cyclone physics is still rather approximate in the statistical downscaling approaches like Localized Constructed Analogs due to their relatively coarse resolution (Strachan et al. 2013). Dynamical downscaling techniques that use process-based models or regional climate models and are driven by GCM projections could provide better opportunities to characterize the dependence of flooding on tropical cyclones (Knutson et al. 2015; Gori et al. 2020).

Our integrated modeling framework enables a quantitative (and approximate) assessment of flood hazards and exposure of cities and boroughs for projected future climatic conditions. Traditionally, flood hazard estimates are only produced under existing conditions using historic flood records to characterize current risks (FEMA 2019). By using GCM projections, we sample key effects of changing climatic conditions to estimate how flood hazards and exposure can change over the decades across cities and boroughs in Pennsylvania. Specifically, we demonstrate a method that leverages an interconnected system of data, models, and analyses to provide nonstationary riverine flood hazard information from local- to regional-scale. Climate-informed flood hazard projections provide a more complete picture of decision-relevant information for the management of flood risks.

*c. Flood risk from future climatic conditions*

The intersection of flood hazards and exposure informs flood-risk assessments (Arnell and Gosling 2014; Hirabayashi et al. 2013; Ruhi et al. 2018; Wing et al. 2018). Flood risk increases with increasing hazards and exposure. We find a strong correlation ($R > 0.90$) between the projected flood hazards and exposure for boroughs (Fig. 5a) and cities (Fig. 5b). Cities with higher flood hazard tend to result in higher exposure, with cities in the Lehigh Valley such as Bethlehem being an exception (Fig 4a-b and Fig. 5b). For the city of Bethlehem, flood hazard of approximately 40 % is projected to result in about 75% of exposure. Bethlehem is one of the most populous cities in Pennsylvania, with development mostly concentrated on both sides of the Lehigh River and Monocacy Creek, which flow through the city. Flood risk in Bethlehem is driven by the compounding effects of floods in the Lehigh River and/or Monocacy Creek. The Lehigh River at Bethlehem has recorded 15 flood events since 1902, while the Lehigh River Valley itself has experienced 129 flood events between 1950 and 2011 (NCEI 2020).

For boroughs (Fig. 5a), the relation between hazards and exposure generally shifts above the one-to-one line, indicating higher exposure for a particular hazard. Increasing exposure with projected flood hazards further suggests that development in the boroughs tends to be concentrated in flood-prone areas. Thus, boroughs may be particularly impacted by the projected increase in flood hazards. This kind of information can be of potential use for borough planning authorities that review regulatory changes in flood hazard zones. Such changes could include, for example, redirecting land development, revisiting flood insurance requirements and/or reducing potential risk by promoting adaptation measures in potential risk zones (Pralle 2019; UNISDR 2015).

Cities and boroughs with highest flood hazards and exposure in current climatic conditions are projected to be roughly the same from future climatic conditions (Fig. 6). In Fig. 6, we show the rank correlation for both flood hazards and exposure between the years 2017 and 2099. The rank correlation for boroughs is shown in Figs. 6(a)-(b) and cities in Figs. 6(c)-(d). We find a strong rank correlation ($R$>0.95) for both flood hazards and exposure between years 2017 and 2099. This implies that cities and boroughs with high flood hazard and exposure in 2017 are projected to be approximately in the same overall situation in 2099 (Fig. 6). However, the frequency distribution for 2099 indicates a heavy tail for both hazards and exposure (see insets in Fig. 6). Based on the frequency distribution of boroughs and cities for flood hazards and exposure in 2017 and 2099 (Fig. 6), flood risk is projected to increase in 2099 compared to 2017. Also, the number of cities and boroughs with high hazards and exposure is projected to increase in 2099. For instance, the number of boroughs with hazards greater than 60% is projected to increase from 96 in 2017 to 170 in 2099 (Fig. 6a), and in terms of exposure, the number of boroughs is projected to increase from 175 in 2017 to 259 in 2099 (Fig. 6b). Moreover, there is no city with exposure greater than 80% in 2017 (refer inset in Fig. 6d); however, there is one city (Sunbury) with exposure projected to be greater than 80% in 2099.

Our results suggest that flood hazards are increasing under the considered future climate conditions. Changing flood hazards may call for changes in flood-mitigation and -adaptation strategies, including floodplain restoration and protection, climate-informed engineering standards, and the use of structural and non-structural measures (Moel et al. 2015).

*d. Uncertainty quantification in flood inundation projections*

We use a case study to quantify the contribution of key sources of uncertainties to the projected flood inundation extents. We focus on Selinsgrove, a riverine community along the main stem of the Susquehanna River, with a total area of 4.92 km$^2$ and a population of 5,654 in 2010 (US Census Bureau 2010). Fig. 7 compares the 100-yr flood inundation extents in the borough of Selinsgrove based on the stationary flood peak estimate from FEMA, nonstationary flood peak estimate in 2017, and nonstationary flood peak projection for years 2060 and 2099. We find that the FEMA 100-yr flood extent is less than the projected flood hazard under nonstationary conditions in 2017 derived in this study (Fig. 7). Maybe more importantly, the projected flood hazards increase with future projections (Fig. 7).

We quantify both the individual and combined sources of uncertainty in different stages of the flood inundation projection chain, including the climate, hydrologic and hydraulic stage. Fig. 8 summarizes the uncertainty quantification results for the 100-yr flood inundation extent in year 2099 using two uncertainty measures: range (Fig. 8a) and standard deviation (Fig. 8b). In Fig. 8, the percentage in the bracket under stage uncertainty indicates the proportion of the uncertainty of each stage contributed to the total uncertainty; while the percentage in the bracket under cumulative uncertainty indicates the proportion of the cumulative uncertainty up to a particular stage.

The total range of projected flood inundation extent in Selinsgrove suggested by this analysis is 0.37 km$^2$, with an inundation extent varying from 2.33 to 2.70 km$^2$ (Fig. 8a). The range in flood inundation extent by just considering the sampling uncertainties in climate model outputs is 0.23 km$^2$. Cumulative uncertainty in the hydrologic stage is 0.33 km$^2$. The cumulative uncertainty in the hydrologic stage is the range of flood inundation extent that considers both climate and hydrologic uncertainties, while keeping the hydraulic stage constant. Thus, the hydrologic stage alone contributes 0.1 km² uncertainty in flood inundation extent. The difference between the total uncertainty in the projected flood inundation extents and cumulative uncertainty in hydrologic stage provides the hydraulic stage uncertainty. The range of the projected flood inundation extents due to the hydraulic stage alone is 0.04 km².

We analyze the uncertainty decomposition results using two measures - range and standard deviation - and arrive at broadly comparable results (Fig. 8). The effects of the considered climate uncertainty dominate the effects of hydrologic and hydraulic uncertainties for both metrics. For instance, when we use the range of the projected flood inundation extents as the uncertainty measure above 60% of the total uncertainty is contributed by the climate model outputs. Thus, reducing the uncertainty in projections of future climate would make a substantial contribution to reducing overall uncertainty in flood inundation projections. The hydrologic stage comprises the second largest uncertainty source, which accounts for more than 25% of the total uncertainty. The combined contribution of climate and hydrologic uncertainty is about 90% of the total uncertainty. The hydraulic stage alone represents about 10% of the total uncertainty.

Flood inundation projections are uncertain and depend on the magnitudes and interactions of the uncertainties in the different stages (e.g. climate, hydrologic and hydraulic) of the projection chain. Prior studies addressing sources of uncertainty in flood projections have focused on quantifying the climate and hydrologic uncertainty. These studies provide important insights but are often silent on the relative contribution of hydraulic uncertainty in the flood-inundation projection chain (Bosshard et al. 2013; Qi et al. 2016; Kim et al. 2019). Our results indicate that all of the investigated uncertainty sources (i.e. climate, hydrologic and hydraulic) make sizeable contributions. Indeed, both uncertainty measures indicate that the hydraulic uncertainty has a non-negligible influence on the projected flood inundation extent. Neglecting the contribution of hydraulic uncertainty in climate change impact assessment can overestimate the relative importance of climate and hydrologic uncertainties.

## 4. Summary and Conclusions

We design, implement, and test an integrated framework to assess regional riverine flood inundation risks in a changing climate. Regional flood inundation modeling can inform the assessment of flood risk across localities and can identify potential risk hotspots. We focus on a case study for the U.S. state of Pennsylvania where many communities have experienced an increase in the frequency and magnitude of floods and associated damages (Shortle et al. 2020). Although the case study focuses on Pennsylvania, the framework is general and can be applied to other regions.

The proposed framework links climate model outputs, a hydrologic model, a statistical model and a hydraulic model. In this implementation, we use downscaled climate projections for precipitation and temperature from CMIP5 models to force the National Oceanic and Atmospheric Administration's (NOAA) Hydrology Laboratory-Research Distributed Hydrologic Model (HL-RDHM) and produce streamflow projections. We fit the Generalized Pareto distribution (GPD) to the projected streamflow and compute extreme flows with a 100-yr return period. We use the 100-yr flood peak as a boundary condition for the hydraulic model LISFLOOD-FP and to generate flood inundation projections. Finally, the flood inundation projections are used to perform flood hazards and exposure analyses. Our regional flood hazards and exposure assessment for Pennsylvania suggests four main conclusions:

1) Assuming a stationary climate can underestimate regional flood risk.

2) Future flood peak alterations due to climate change are projected to be greater for smaller basins, which capture the spatial scale at which many flood-related infrastructure decisions are made.

3) Most cities and boroughs in Pennsylvania are projected to face higher flood risk under projected future climate conditions. Flood risks are relatively higher in smaller cities and boroughs compared to the largest cities. However, the cities and boroughs with highest flood risk in 2017 are projected to be roughly the same in 2099.

4) Climate uncertainty is the dominant contributor to flood hazard uncertainties. The hydrologic stage is the second largest uncertainty source. We also find that hydraulic uncertainty has an important influence on extreme flood uncertainty. Neglecting the contribution of hydraulic uncertainty to the total uncertainty in climate change impact assessment could lead to overconfident estimates of flood hazard uncertainty.

This study provides an integrated, multi-model and multiscale framework to develop local-to-regional scale flood-risk information in a changing climate. Avenues for future work to overcome limitations in our study include (i) the explicit inclusion of relevant physical infrastructure in the hydrologic and hydraulic model, and (ii) a more comprehensive uncertainty quantification by incorporating additional sources of uncertainties, including different emission and land use cover scenarios, climate and hydrologic models, downscaling approaches, and sets of hydrologic model parameters.

**Acknowledgements**

This work was supported through the Penn State Initiative for Resilient Communities (PSIRC) by a Strategic Plan seed grant from the Penn State Office of the Provost, with co-support from the Center for Climate Risk Management (CLIMA), the Rock Ethics Institute, Penn State Law, and the Hamer Center for Community Design. All errors and opinions are from the authors and do not reflect the funding agencies. We are grateful to Lisa Iulo, James Ross-Golin, Lara Fowler, Skip Wishbone and Irene Schaperdoth for their helpful feedback and advice.

**Competing interests**

The authors declare no competing financial or nonfinancial interests.

**Legal disclaimer**

This is academic research and not designed to be used in actual decision-making. All data and code used for this analysis will be made available upon publication, under the GNU general public open-source license. The results, data, software tools, and other resources related to this work are provided as-is without warranty of any kind, expressed or implied. In no event shall the authors or copyright holders be liable for any claim, damages or other liability in connection with the use of these resources.

# REFERENCES

Aich, V., and Coauthors, 2016: Flood projections within the Niger River Basin under future land use and climate change. *Science of the Total Environment*, 562, 666-677.

Alfonso, L., M. M. Mukolwe, and G. Di Baldassarre, 2016: Probabilistic flood maps to support decision-making: Mapping the value of information. *Water Resources Research*, 52, 1026-1043.

Alfieri, L., L. Feyen, F. Dottori, and A. Bianchi, 2015: Ensemble flood risk assessment in Europe under high end climate scenarios. *Global Environmental Change*, 35, 199-212.

Alfieri, L., B. Bisselink, F. Dottori, G. Naumann, A. Roo, P. Salamon, K. Wyser, and L. Feyen, 2017: Global projections of river flood risk in a warmer world. *Earth's Future*, 5, 171-182.

Armstrong, W. H., M. J. Collins, and N. P. Snyder, 2014: Hydroclimatic flood trends in the northeastern United States and linkages with large-scale atmospheric circulation patterns. *Hydrological Sciences Journal*, 59, 1636-1655.

Arnell, N. W., and S. N. Gosling, 2014: The impacts of climate change on river flood risk at the global scale. *Climatic Change*, 1–15, doi:10.1007/s10584-014-1084-5.

Bales, J. D., and C. R. Wagner, 2009: Sources of uncertainty in flood inundation maps. *J. Flood Risk Manage.*, 2, 139–147.

Bates, P., and Coauthors, 2013. Lisflood-FP. User manual. School of Geographical Sciences, University of Bristol. Bristol, UK.

Bosshard, T., M. Carambia, K. Goergen, S. Kotlarski, P. Krahe, M. Zappa, and C. Schär, 2013: Quantifying uncertainty sources in an ensemble of hydrological climate-impact projections, *Water Resources Research*, 49(3), 1523-1536.

Bracken, C., K. D. Holman, B. Rajagopalan, and H. Moradkhani, 2018: A Bayesian hierarchical approach to multivariate nonstationary hydrologic frequency analysis. *Water Resources Research*, 54, 243-255.

Chaplin, J. J., 2005: Development of regional curves relating bankfull-channel geometry and discharge to drainage area for streams in Pennsylvania and selected areas of Maryland (p. 34). US Department of the Interior, US Geological Survey, http://pubs.usgs.gov/sir/2005/5147/ .

Chen, J., F. P. Brissette, A. Poulin, and R. Leconte, 2011: Overall uncertainty study of the hydrological impacts of climate change for a Canadian watershed. *Water Resour. Res*., 47, W12509, doi:10.1029/2011WR010602.

Cheng, C., Y. E. Yang, R. Ryan, Q. Yu, and E. Brabec, 2017: Assessing climate change-induced flooding mitigation for adaptation in Boston's Charles River watershed, USA. *Landscape and Urban Planning*, 167, 25-36.

Cinotto, P. J., 2003: Development of regional curves of bankfull-channel geometry and discharge for streams in the non-urban, Piedmont Physiographic Province, Pennsylvania and Maryland (No. 3). US Department of the Interior, US Geological Survey, http://pubs.er.usgs.gov/publication/wri034014 .

Clune, J. W., J. J. Chaplin, and K. E. White, 2018: Comparison of regression relations of bankfull discharge and channel geometry for the glaciated and nonglaciated settings of Pennsylvania and southern New York (ver. 1.1, July 2020): U.S. Geological Survey Scientific Investigations Report 2018–5066, 20 p., https://doi.org/10.3133/sir20185066.

Coles, S. G., 2001: An Introduction to Statistical Modeling of Extreme Values. Springer, 225 pp.

Collet, L., L. Beevers, and M. D. Stewart, 2018: Decision-Making and Flood Risk Uncertainty: Statistical Data Set Analysis for Flood Risk Assessment. *Water Resources Research,* 54, 7291-7308.

Demaria, E. M., R. N. Palmer, and J. K. Roundy, 2016: Regional climate change projections of streamflow characteristics in the Northeast and Midwest US. *Journal of Hydrology: Regional Studies,* 5, 309-323.

De Moel, H., B. Jongman, H. Kreibich, B. Merz, E. Penning-Rowsell, and P. J. Ward, 2015: Flood risk assessments at different spatial scales. *Mitigation and Adaptation Strategies for Global Change*, 20, 865-890.

Dibike, Y. B., and P. Coulibaly, 2005: Hydrologic impact of climate change in the Saguenay watershed: comparison of downscaling methods and hydrologic models. *Journal of hydrology*, 307, 145-163.

Dobler, C., G. Bürger, and J. Stötter, 2012: Assessment of climate change impacts on flood hazard potential in the Alpine Lech watershed. *Journal of Hydrology*, 460, 29-39.

Brooks, L.T., 2005: Flood of September 18-19, 2004 in the Upper Delaware River Basin, New York: U.S. *Geological Survey Open-File Report* 2005-1166, 123 p.

DRBC, 2006: Delaware River Basin Commission, Flooding events in the Delaware River basin. Retrieved on July 2019, https://www.state.nj.us/drbc/hydrological/flood/drb-flood-events.html.

England, J. F. Jr., T. A. Cohn, B. A. Faber, J. R. Stedinger, W. O. Thomas, Jr., A. G. Veilleux, J. E. Kiang, and R. R. Jr. Mason, 2018: "Guidelines for determining flood flow frequency—Bulletin 17C." Chap. B5 in USGS Techniques and Methods, Book 4. Reston, VA: USGS.

Feyen, L., R. Dankers, K. Bódis, P. Salamon, and J. I. Barredo, 2012: Fluvial flood risk in Europe in present and future climates. *Climatic change*, 112, 47-62.

FEMA, 2019: Flood Insurance Rate map (FIRM), assessed July 2019, https://www.fema.gov/flood-insurance-rate-map-firm.

FCSMA, 2020: Flood Control Sunbury Municipal Authority, Asses July 2020, http://sunburyfloodcontrol.com/flood-history.html

FSF, 2020: The First Street Foundation, The First National Flood Risk Assessment: assessed July 2020, https://assets.firststreet.org/uploads/2020/06/first_street_foundation__first_national_flood_risk_assessment.pdf

Gangrade, S., S. C. Kao, and R. A. McManamay, 2020: Multi-model Hydroclimate projections for the Alabama-coosa-tallapoosa River Basin in the Southeastern United States. *Scientific reports*, 10(1), 1-12.

Gitro, C. M., M. S. Evan and R. H. Grumm, 2014: Two Major Heavy Rain/Flood Events in the Mid-Atlantic: June 2006 and September 2011. *Journal of Operational Meteorology*, 2(13).

Gomez, M., S. Sharma, S. Reed and A. Mejia, 2019: Skill of ensemble flood inundation forecasts at short- to medium-range timescales. Journal of Hydrology, 568, 207–220.

Gori, A., N. Lin and J. Smith, 2020: Assessing Compound Flooding From Landfalling Tropical Cyclones on the North Carolina Coast. *Water Resources Research*, 56(4), e2019WR026788.

Hayhoe, K., Wuebbles, D.J., Easterling, D.R., Fahey, D.W., Doherty, S., Kossin, J., Sweet, W., Vose, R., Wehner, M., 2018: Our Changing Climate, in: Impacts, Risks, and Adaptation in the United States: *Fourth National Climate Assessment*, Volume II. pp. 72–144. https://doi.org/10.7930/NCA4.2018.CH2

Hattermann, F. F., and Coauthors, 2017: Cross-scale intercomparison of climate change impacts simulated by regional and global hydrological models in eleven large river basins. *Climatic Change*, 141, 561-576.

Hirabayashi, Y., R. Mahendran, S. Koirala, L. Konoshima, D. Yamazaki, S. Watanabe, H. Kim, and S. Kanaes, 2013: Global flood risk under climate change. *Nature Climate Change*, 3, 816–821.

Homer, C.H., J.A. Fry, and C.A. Barnes, 2012: The National Land Cover Database, U.S. Geological Survey Fact Sheet 2012-3020, 4 p.

Huang, S., F. F. Hattermann, V. Krysanova, and A. Bronstert, 2013: Projections of climate change impacts on river flood conditions in Germany by combining three different RCMs with a regional eco-hydrological model. *Climatic change*, 116, 631-663.

Jongman, B., and Coauthors, 2015: Declining vulnerability to river floods and the global benefits of adaptation. *Proc. Natl Acad. Sci.,* 112, E2271–E2280.

Jovanovic, T., A. Mejía, H. Gall and J. Gironás, 2016: Effect of urbanization on the long-term persistence of streamflow records. Physica A: Statistical Mechanics and its Applications, 447, 208-221. doi:http://dx.doi.org/10.1016/j.physa.2015.12.024.

Jovanovic, T., F. Sun, T. Mahjabin and A. Mejia, 2018: Disentangling the effects of climate and urban growth on streamflow for sustainable urban development: A stochastic approach to flow regime attribution. Landscape and Urban Planning, 177, 160-170. doi:https://doi.org/10.1016/j.landurbplan.2018.05.009

Judi, D. R., C.L. Rakowski, S.R. Waichler, Y. Feng, and M.S. Wigmosta, 2018: Integrated modeling approach for the development of climate-informed, actionable information. *Water*, 10(6), 775.

Kang, H., V. Sridhar, 2018: Assessment of Future Drought Conditions in the Chesapeake Bay Watershed. *JAWRA J. Am. Water Resour. Assoc,* 54, 160–183. https://doi.org/10.1111/1752-1688.12600.

Keller, K., C. Helgeson, and V. Srikrishnan, 2020: Climate Risk Management. Annual Review of Earth and Planetary Sciences, 49, accepted. Retrieved from https://personal.ems.psu.edu/~kzk10/Keller_et_al_Climate Risk Management_Preprint_2020docx.pdf

Kim, Y., I. Ohn, J. K. Lee, and Y. O. Kim, 2019: Generalizing uncertainty decomposition theory in climate change impact assessments, Journal of Hydrology X, 3, 100024.

Knutson, T. R., J. J. Sirutis, M. Zhao, R. E. Tuleya, M. Bender, G. A. Vecchi, G. Villarini, and D. Chavas, 2015: Global projections of intense tropical cyclone activity for the late 21st century from dynamical downscaling of CMIP5/RCP4.5 scenarios. *J. Climate*, **28**, 7203–7224, doi:10.1175/JCLI-D-15-0129.1.

Koren, V., S. Reed, M. Smith, Z. Zhang, and D.-J. Seo, 2004: Hydrology Laboratory Research Modeling System (HL-RMS) of the US National Weather Service. J. Hydrol., 291, 297–318, doi: 10.1016/j.jhydrol.2003.12.039.

Koutsoyiannis, D., 2006: Nonstationarity versus scaling in hydrology, *Journal of Hydrology*, 324(1-4), 239-254.

Kundzewicz, Z. W., D. L. T. Hegger, P. Matczak, and P. P. J. Driessen, 2018: Opinion: Flood-risk reduction: Structural measures and diverse strategies. *Proceedings of the National Academy of Sciences*, 115, 12321-12325.

Kuzmin, V. A., 2009: Algorithms of automatic calibration of multiparameter models used in operational systems of flash flood forecasting. Russ. Meteor. Hydrol., 34, 473–481, doi:10.3103/ S1068373909070073.

Kuzmin, V. A., D.-J. Seo, and V. Koren, 2008: Fast and efficient optimization of hydrologic model parameters using a priori estimates and stepwise line search. J. Hydrol., 353, 109–128, doi:10.1016/ j.jhydrol.2008.02.001.

Lang, M., T. Ouarda, B. Bobée, 1999: Towards operational guidelines for over-threshold modeling. *Journal of hydrology* 225(3-4), 103-117.

Lin, S., C. Jing, N. A. Coles, V. Chaplot, N. J. Moore, and J. Wu, 2013: Evaluating DEM source and resolution uncertainties in the Soil and Water Assessment Tool. *Stochastic environmental research and risk assessment*, 27, 209-221.

Iulo, L., and Coauthors, 2020: Establishing Priorities for Pennsylvania Community Flood Resilience. PSIRC. Retrieved from https://www.psirc.psu.edu/misc/Iulo_etal_2020.Establishing_Priorities_White_Paper.pdf, accessed Sept 17 2020.

McCuen, R. H., and W. M. Snyder, 1975: A proposed index for comparing hydrographs. Water Resour. Res., 11, 1021–1024, doi:10.1029/WR011i006p01021.

Meinshausen, M and coauthors ,2011: The RCP greenhouse gas concentrations and their extensions from 1765 to 2300. *Climatic change*, 109(1-2), 213.

Merwade, V. M., D. R. Maidment, and J. A. Goff, 2006: Anisotropic considerations while interpolating river channel bathymetry. *J. Hydrol.*, 331(3–4), 731–741.10.1016/j.jhydrol.2006.06.018.

Merwade V., F. Olivera, M. Arabi, and & S. Edleman, 2008: Uncertainty in flood inundation mapping: current issues and future directions. *J. Hydrol. Eng.*, 13, 608–620.

Nash, J. E., and J. V. Sutcliffe, 1970: River flow forecasting through conceptual models part I – A discussion of principles, J. Hydrol., 10, 282–290.

NCEI, National Center for Environmental Information Storm Events Database, 2020: Search results for all U.S. states and areas, https://www.ncdc.noaa.gov/stormevents/.

Neal, J., I. Villanueva, N. Wright, T. Willis, T. Fewtrell, and P. D. Bates, 2012: How much physical complexity is needed to model flood inundation?, Hydrol. Process., 26, 2264–2282, https://doi.org/10.1002/hyp.8339.

Newlin, J. T., and B. R. Hayes, 2015: Hydraulic modeling of glacial dam-break floods on the West Branch of the Susquehanna River, Pennsylvania. *Earth and Space Science*, 2, 229-243.

Pappenberger, F., K. Beven, M. Horritt, and S. Blazkova, 2005: "Uncertainty in the calibration of effective roughness parameters in HEC-RAS using inundation and downstream level observations." J. Hydrol., 302(1–4), 46–69.10.1016/j.jhydrol.2004.06.036.

Pappenberger, F., P. Matgen, K. J. Beven, J. B. Henry, and L. Pfister, 2006: Influence of uncertain boundary conditions and model structure on flood inundation predictions. *Advances in water resources*, 29, 1430-1449.

Papoulis, A., and S. U. Pillai, 2002: Probability, random variables, and stochastic processes. Tata McGraw-Hill Education.

Pennsylvania Emergency Management Agency (PEMA), 2017: Summary of Commonwealth Vulnerability Analysis. Retrieve from http://www.pema.pa.gov/planningandpreparedness/communityandstateplanning/Pages/HazardVulnerability.aspx#.WYDWb_nyvcs.

Pierce, D. W., D. R. Cayan, and B. L. Thrasher, 2014: Statistical downscaling using localized constructed analogs (LOCA). *Journal of Hydrometeorology*, 15(6), 2558-2585.

Prat, O. P., and B. R. Nelson, 2015: Evaluation of precipitation estimates over CONUS derived from satellite, radar, and rain gauge data sets at daily to annual scales (2002–2012). *Hydrol. Earth Syst. Sci.*, 19, 2037–2056, doi:10.5194/hess-19-2037-2015

Pralle, S., 2019: Drawing lines: FEMA and the politics of mapping flood zones. *Climatic change*, 152, 227-237.

Qi, W., C. Zhang, G. Fu, H. Zhou, and J. Liu, 2016: Quantifying uncertainties in extreme flood predictions under climate change for a medium-sized basin in Northeastern China. *Journal of Hydrometeorology*, 17, 3099-3112.

Reed, S., V. Koren, M. Smith, Z. Zhang, F. Moreda, D.-J. Seo, and D. Participants, 2004: Overall Distributed Model Intercomparison Project results. J. Hydrol., 298, 27–60, doi:10.1016/ j.jhydrol.2004.03.031.

Roland, M.A., and S. A. Hoffman, 2011: Development of flood-inundation maps for the West Branch Susquehanna River near the Borough of Jersey Shore, Lycoming County, Pennsylvania: *U.S. Geological Survey Scientific Investigations Report* 2010–5057, 9 p.

Ruhi, A., M. L. Messager, and J. D. Olden, 2018: Tracking the pulse of the Earth's fresh waters, *Nature Sustainability*, 1(4), 198-203.3

Sagarika, S., A. Kalra, and S. Ahmad, 2014: Evaluating the effect of persistence on long-term trends and analyzing step changes in streamflows of the continental United States. *Journal of Hydrology*, 517, 36-53.

Sanders, B. F., and Coauthors, 2020: Collaborative modeling with fine-resolution data enhances flood awareness, minimizes differences in flood perception, and produces actionable flood maps. *Earth's Future*, 7, e2019EF001391.

Sharma, S., R. Siddique, S. Reed, P. Ahnert, P. A. Mendoza Zúñiga, and A. Mejia, 2018: Relative effects of statistical preprocessing and postprocessing on a regional hydrological ensemble prediction system. Hydrol. Earth Syst. Sci., 22, 1831– 1849, https://doi.org/10.5194/hess-22-1831-2018

Shortle, J., and Couthors, 2015: Pennsylvania Climate Impacts Assessment-Update. Pennsylvania Department of Environmental Protection.

Shortle, J., and Couthors, 2020: Pennsylvania Climate Impacts Assessment-Update. Pennsylvania Climate Impacts Assessment - Update. Pennsylvania Department of Environmental Protection.

Siddique, R., and A. Mejia, 2017: Ensemble Streamflow Forecasting across the U.S. Mid-Atlantic Region with a Distributed Hydrological Model Forced by GEFS Reforecasts. Journal of Hydrometeorology, 18, 1905–1928. https://doi.org/10.1175/JHM-D-16-0243.1

Smith, J.A., M. L. Baeck, G. Villarini, and W. F. Krajewski, 2010: The hydrology and hydrometeorology of flooding in the Delaware River basin. J. Hydrometeor., 11, 841–869.

Smith, J. A., G. Villarini and M. L. Baeck, 2011: Mixture distributions and the hydroclimatology of extreme rainfall and flooding in the eastern United States. J. Hydrometeor., 12, 294–309, doi:10.1175/2010JHM1242.1.

Smith M. B., and Coauthors, 2012: The Distributed Model Intercomparison Project—Phase 2: Motivation and design of the Oklahoma experiments. J. Hydrol., 418–419, 3–16, doi:10.1016/j.jhydrol.2011.08.055.

Sorribas, M. V., and Coauthors, 2016: Projections of climate change effects on discharge and inundation in the Amazon basin. *Climatic change*, 136, 555-570.

Sriver, R. L., C. E. Forest, and K. Keller, 2015: Effects of initial conditions uncertainty on regional climate variability: An analysis using a low-resolution CESM ensemble. *Geophysical Research Letters*, 42(13), 5468–5476. https://doi.org/10.1002/2015gl064546.

Srikrishnan, V., R. Alley, and K. Keller, 2019: Investing in science to improve climate risk management, Eos, 100, https://doi.org/10.1029/2019EO131077. Published on 16 August 2019.

Stephenson, A., and J. Tawn, 2004: Bayesian inference for extremes: accounting for the three extremal types. Extremes 7, 291–307.

Strachan, J., P. L. Vidale, K. Hodges, M. Roberts, and M.-E. Demory, 2013: Investigating global tropical cyclone activity with a hierarchy of AGCMs: The role of model resolution. J. Climate, 26, 133–152, doi:10.1175/JCLI-D-12-00012.1.

Suro, T. P., G. D. Firda and C. O. Szabo, 2009: Flood of June 26–29, 2006, Mohawk, Delaware, and Susquehanna River Basins, New York: U.S. Geological Survey Open-File Report 2009–1063, 354p. Available online at http://pubs.usgs.gov/ofr/2009/1063.

Taylor, K. E., R. J. Stouffer, and G. A. Meehl, 2012: An Overview of CMIP5 and the Experiment Design. Bull. Am. Meteorol. Soc., 93, 485–498, https://doi.org/10.1175/bams-d-11-00094.1.

United Nations Office for Disaster Risk Reduction (UNISDR), 2015: Making Development Sustainable: The Future of Disaster Risk Management. Global Assessment Report on Disaster Risk Reduction; www.unisdr.org/we/inform/publications/ 42809

U.S. Department of Energy, Office of Electricity Delivery and Energy Reliability, "State of Pennsylvania Energy Sector Risk Profile" (USDOE), 2015: https://www.energy.gov/sites/prod/files/2015/05/f22/PA-Energy%20Sector%20Risk%20Profile.pdf

US Census Bureau, 2010: 2010 Census, Profile of General Population and Housing Characteristics: 2010 Demographic Profile Data, Table DP-1.

Varis, O., T. Kajander, and R. Lemmelä, 2004: Climate and water: from climate models to water resources management and vice versa. *Climatic Change*, 66, 321-344.

Vetter T., and Coauthors, 2017: Evaluation of sources of uncertainty in projected hydrological changes under climate change in 12 large-scale river basins. *Clim. Change*, 141, 419–33

Villarini, G., J. A. Smith, F. Serinaldi, J. Bales, P. D. Bates, and W.F. Krajewski, 2009: Flood frequency analysis for nonstationary annual peak records in an urban drainage basin. *Advances in water resources*, 32, 1255-1266.

Ward, P. J., and Coauthors, 2015: Usefulness and limitations of global flood risk models. *Nature Clim. Change*, 5, 712–715.

Winsemius, H. C., and Coauthors, 2016: Global drivers of future River flood risk. *Nat. Clim. Change*, 6, 381–5.

Wing O. E. J., P. D. Bates, A. M. Smith, C. C. Sampson, K. A. Johnson, J. Fargione, and P. Morefield, 2018: Estimates of present and future flood risk in the conterminous United States. *Environ. Res. Lett.* 13 034023

Wu, C., G. Huang, H. Yu, Z. Chen, and J. Man 2014: Impact of climate change on reservoir flood control in the upstream area of the Beijiang River Basin, South China. *Journal of Hydrometeorology*, 15, 2203-2218.

Zarzar, C. M., H. Hosseiny, R. Siddique, M. Gomez, V. Smith, A. Mejia and J. Dyer, 2018: A Hydraulic MultiModel Ensemble Framework for visualizing flood inundation uncertainty. Journal of the American Water Resources Association, 54,807– 819. https://doi.org/10.1111/1752-1688.12656.
36

Zarekarizi, M., V., Srikrishnan, and K. Keller, 2020: Neglecting Uncertainties Leads to Suboptimal Decisions About Home-Owners Flood Risk Management. arXiv preprint arXiv:2001.06457.

Zhang, Z., A. D. Dehoff, R. D. Pody, and J.W. Balay, 2010: Detection of streamflow change in the Susquehanna River Basin. *Water resources management*, 24, 1947-1964.

**Table 1**. The thirteen models of the Fifth Coupled Model Intercomparison Project Phase 5 (CMIP5) used in this study.

| Modeling Center (or Group) | Institute ID | Model Name |
|---|---|---|
| Beijing Climate Center, China Meteorological Administration | BCC | BCC-CSM1.1 |
| Community Earth System Model Contributors | NSF-DOE-NCAR | CESM1-CAM5 |
| Canadian Centre for Climate Modelling and Analysis | CCCMA | CanESM2 |
| EC-EARTH consortium | EC-EARTH | EC-EARTH |
| NOAA Geophysical Fluid Dynamics Laboratory | NOAA GFDL | GFDL-ESM2M |
| NASA Goddard Institute for Space Studies | NASA GISS | GISS-E2-R |
| Met Office Hadley Centre | MOHC | HadGEM2-CC |
| Met Office Hadley Centre | MOHC | HadGEM2-ES |
| Institut Pierre-Simon Laplace | IPSL | IPSL-CM5A-LR |
| Institut Pierre-Simon Laplace | IPSL | IPSL-CM5A-MR |
| Max-Planck-InstitutfürMeteorologie (Max Planck Institute for Meteorology) | MPI-M | MPI-ESM-LR |
| Max-Planck-InstitutfürMeteorologie (Max Planck Institute for Meteorology) | MPI-M | MPI-ESM-MR |
| Institute of Numerical Mathematics | INM | INM-CM4 |

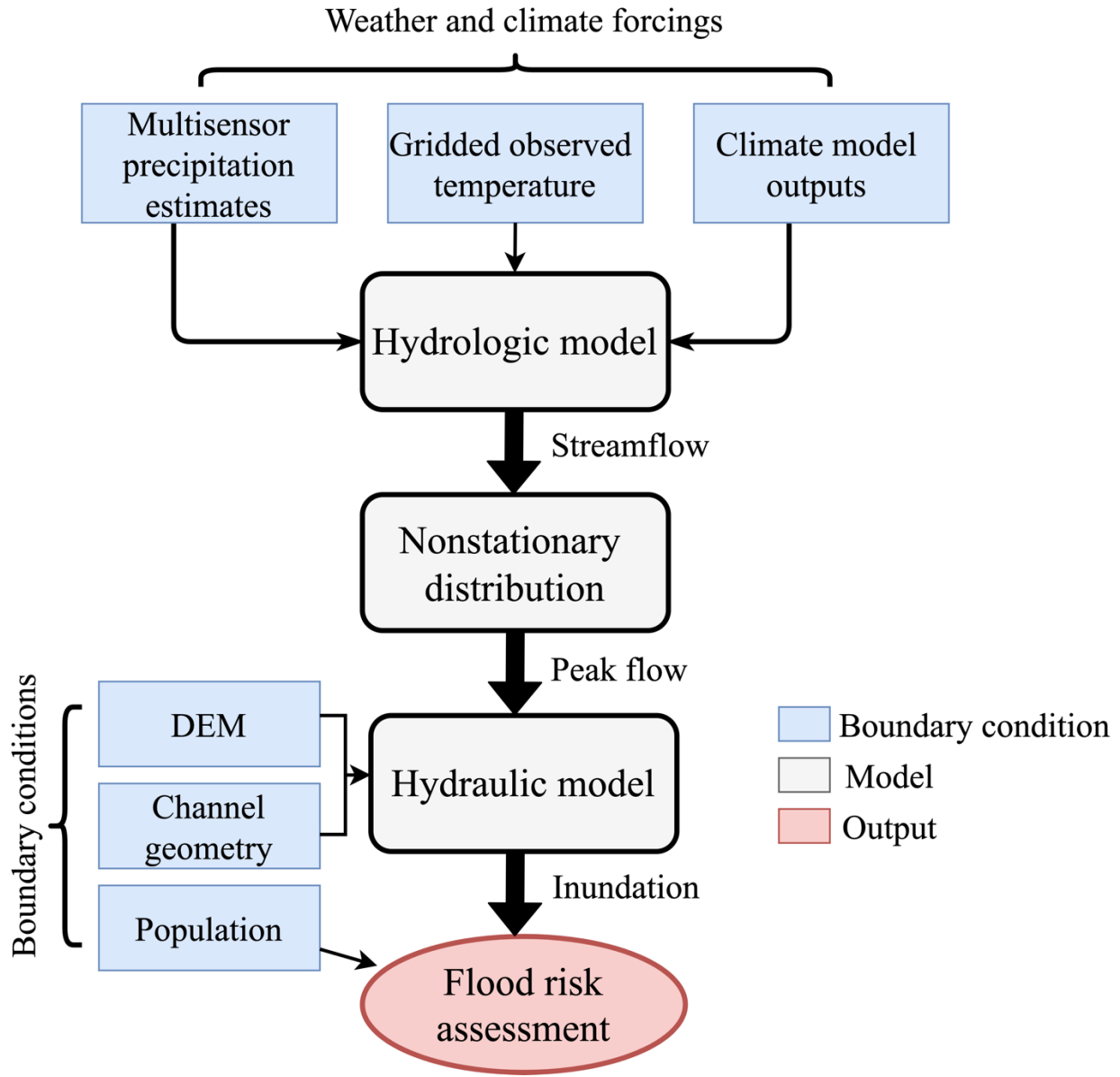

**Figure 1**. Flowchart illustrating the general methodological approach for flood inundation risk projections. The approach starts with the climate model outputs, which are used to drive the hydrologic model and generate streamflow projections. Together with the statistical and hydraulic model, the projected streamflow is then used to map the uncertainty of flood inundation projections for extreme flood events. *DEM= Digital Elevation Model.*

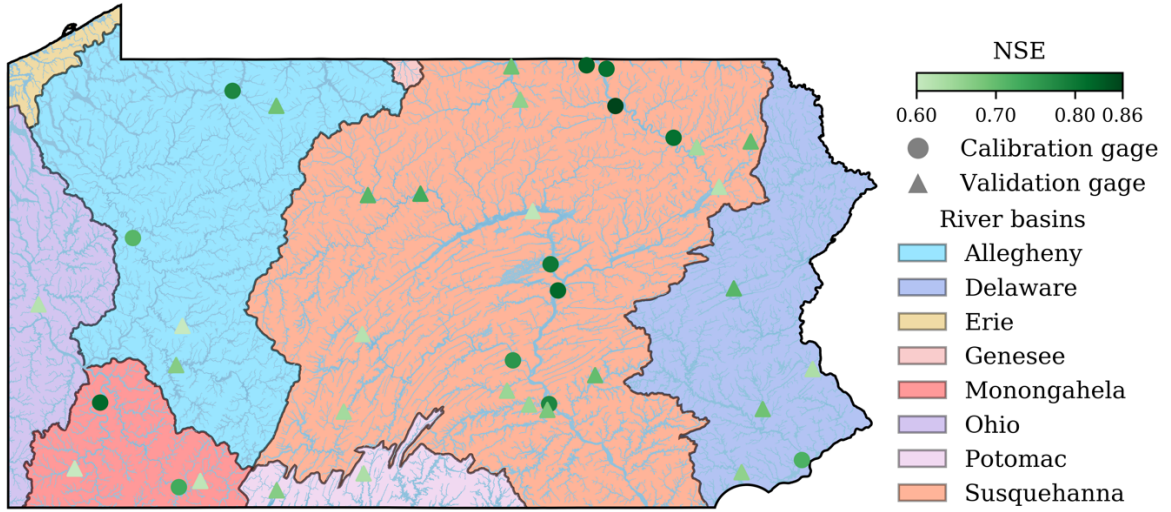

**Figure 2**. Map illustrating the major river basins in Pennsylvania. The map also shows the hydrologic model calibration and validation gage stations, and the corresponding performance of the hydrologic simulations based on the Nash-Sutcliffe Efficiency (*NSE*) index.

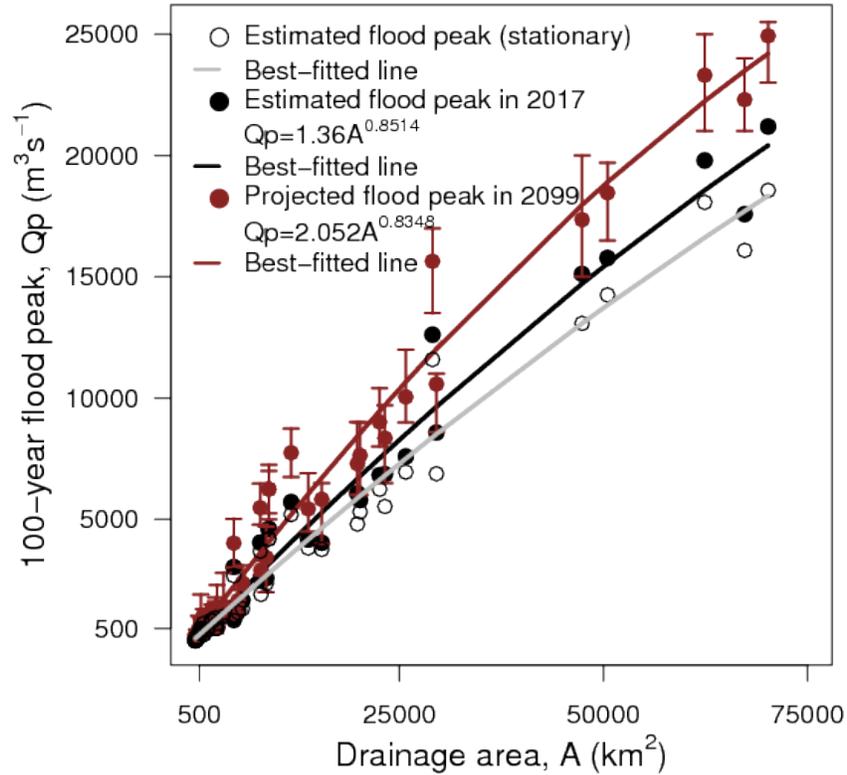

**Figure 3**. Scaling of the 100-yr flood peak with the drainage area for locations in Pennsylvania. The scaling relationship is shown for current (2017) and projected flood peak estimates (2099). Flood peak estimates for year 2017 are based on the historical flood records under the stationary and nonstationary assumption. The flood peaks for 2099 depict the mean flood projections, and the corresponding error bar represents the maximum and minimum projected flood peak. The solid lines represent the best fit line under the stationary assumption (gray line) as well as the nonstationary assumption in 2017 (black line) and 2099 (brown line).

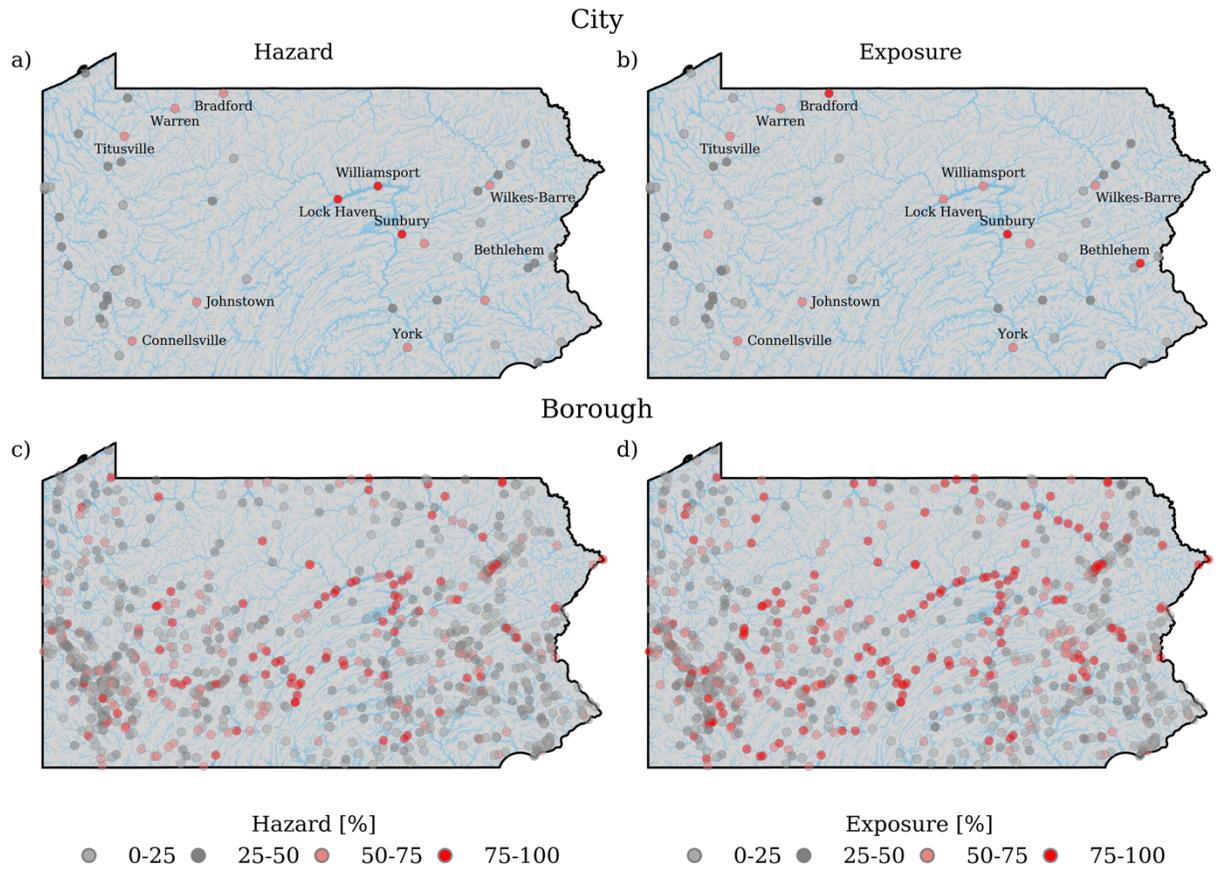

**Figure 4**. Projected flood hazard and exposure in year 2099 for all the (a)-(b) cities and (c)-(d) boroughs in Pennsylvania. Hazard [%] indicates the inundation area standardized to the total area of a city or borough. Exposure [%] indicates the population in the projected flood inundation extents standardized to the total population of a city or borough. There are 959 boroughs and 56 cities in Pennsylvania.

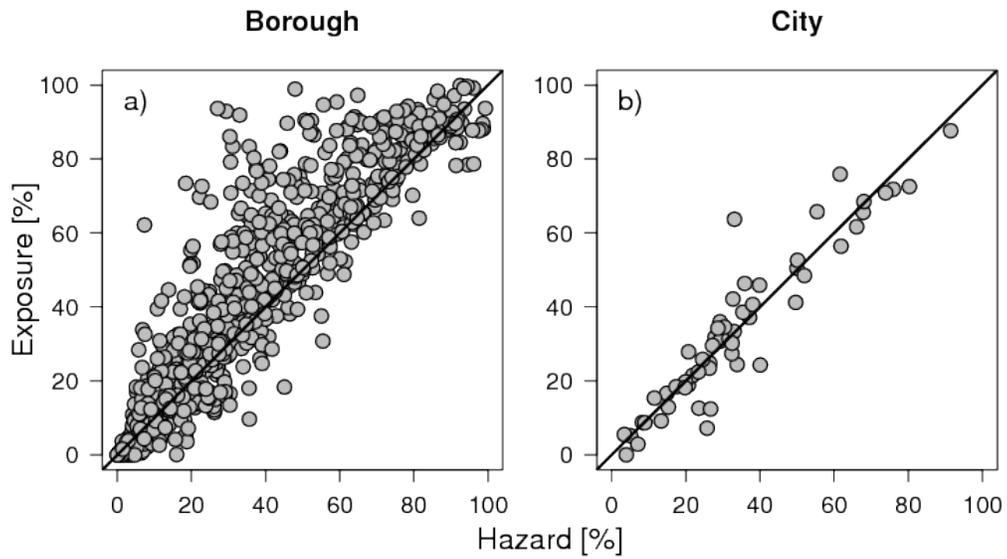

**Figure 5**. Correlation between the projected flood hazard and exposure in year 2099 for a) boroughs and b) cities. The black line represents the 1:1 fit. Hazard [%] indicates the inundation area standardized to the total area of a city or borough. Exposure [%] indicates the population in the projected flood inundation extents standardized to the total population of a city or borough. The value of the correlation coefficient between hazard and exposure is 0.91 for boroughs and 0.94 for cities.

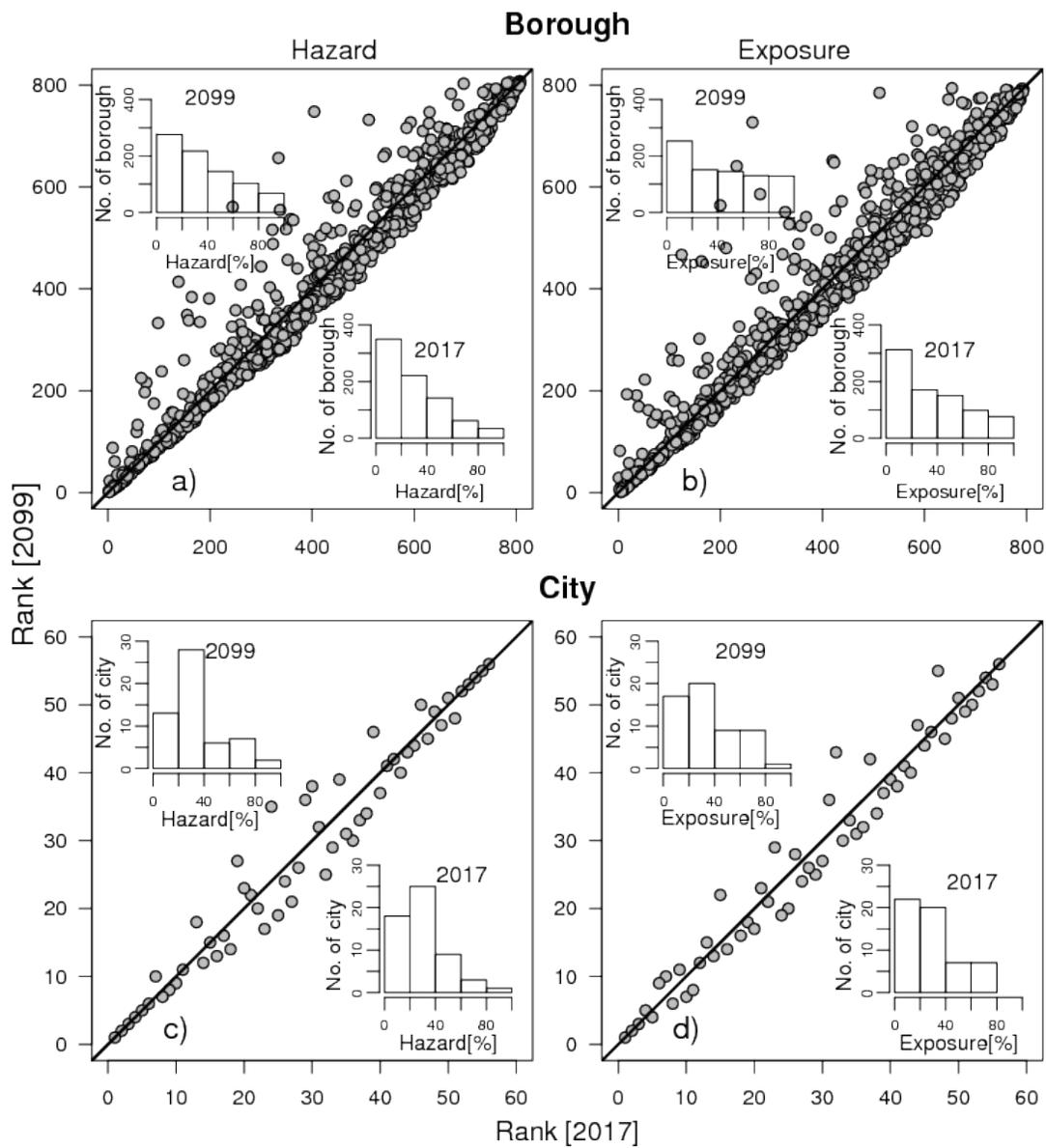

**Figure 6**. Rank correlation between years 2017 and 2099 for the flood hazard and exposure of (a)-(b) boroughs and (c)-(d) cities. The insets show the frequency distribution of standardized flood hazard and exposure for boroughs and cities.

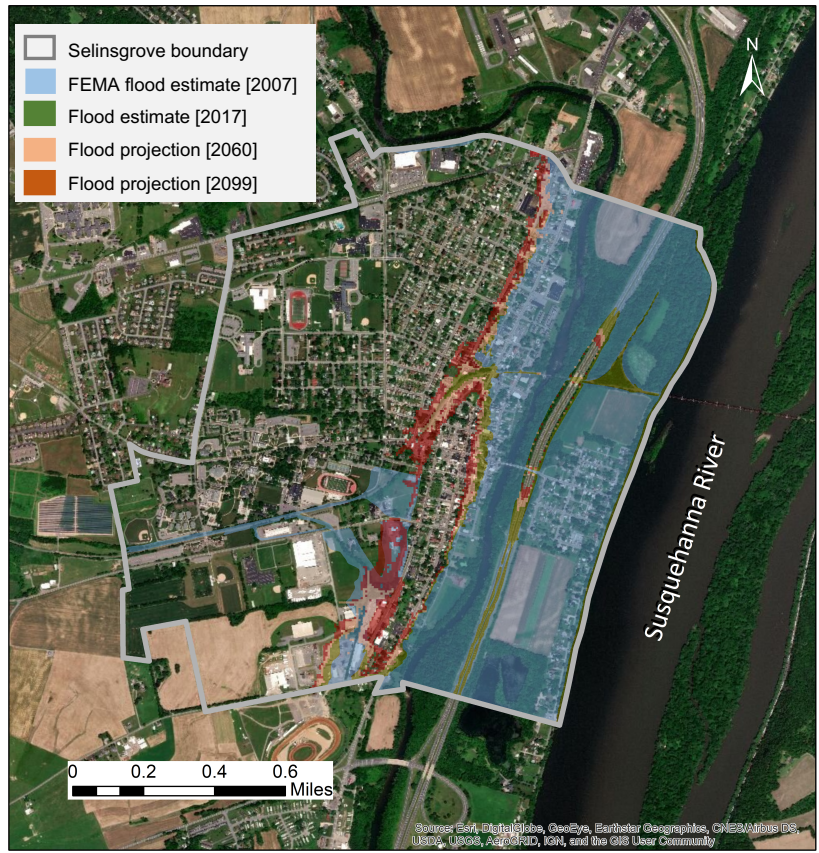

**Figure 7**. The 100-yr flood inundation extents in the borough of Selinsgrove, Pennsylvania, using FEMA stationary flood peak estimates (FEMA, 2019), nonstationary flood peak estimates in year 2017, and nonstationary flood peak projections for years 2060 and 2099. *FEMA*=Federal Emergency Management Agency.

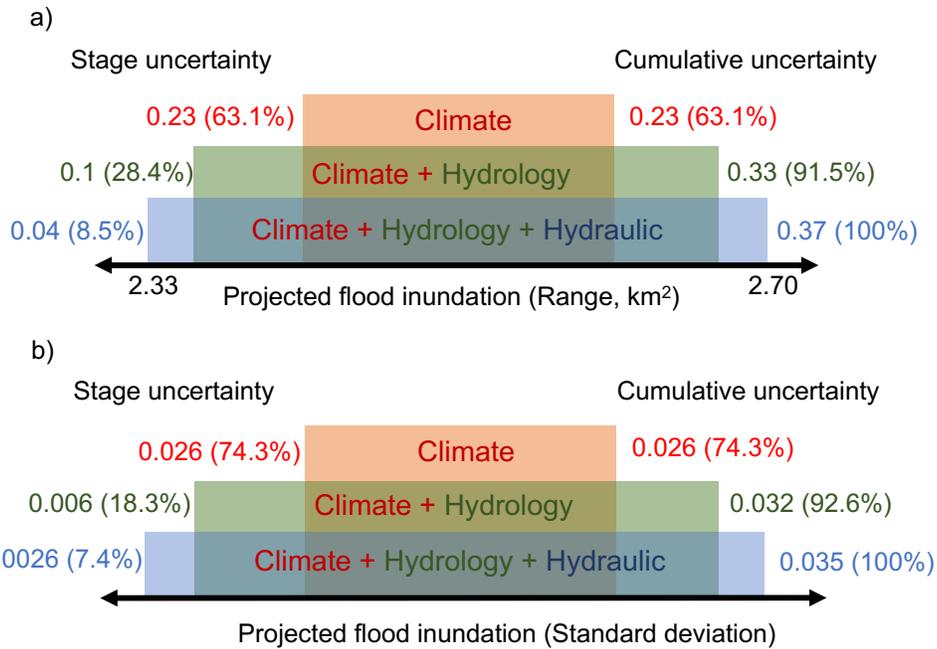

**Figure 8**. Decomposition of the uncertainty in the projected flood inundation extents for year 2099 using two different measures: a) range and b) standard deviation. The individual and combined sources of uncertainties (climate, hydrology and hydraulics) in the flood inundation projection chain are shown. The percentage values in the brackets under stage uncertainty indicate the proportion of the uncertainty that each stage contributes to the total uncertainty. The percentage values in the brackets under cumulative uncertainty indicate the proportion of the cumulative uncertainty up to that particular stage.